\documentclass[10pt,conference]{IEEEtran}
\usepackage[a4paper,
            bindingoffset=0.2in,
            left=0.68in,
            right=0.68in,
            top=0.7in,
            bottom=1.8in,
            footskip=.25in]{geometry}

\usepackage{amsmath,amssymb,amsfonts}
\usepackage{graphicx}
\usepackage{textcomp}
\usepackage{xcolor}
\usepackage{mathptmx}
\usepackage{siunitx}
\usepackage{float}
\usepackage{epstopdf}
\usepackage{graphicx}
\usepackage{caption}
\usepackage{subcaption}
\usepackage{comment}
\usepackage{lipsum}
\usepackage{tikz,cite}
\usepackage{pgfplots}
\usetikzlibrary{decorations.pathreplacing,positioning}
\usepackage{caption,multirow,booktabs}
\usepackage{subcaption}
\usepackage{soul}
\usepackage[multiple]{footmisc}
\makeatletter
\def\ps@IEEEtitlepagestyle{%
  \def\@oddfoot{\mycopyrightnotice}%
}
\def\mycopyrightnotice{%
  \begin{minipage}{\textwidth}
  \centering \scriptsize
  \vspace{10pt}
  \copyright 2025 IEEE. Personal use of this material is permitted. Permission from IEEE must be obtained for all other uses, in any current or future media, including reprinting/republishing this material for advertising or promotional purposes, creating new collective works, for resale or redistribution to servers or lists, or reuse of any copyrighted component of this work in other works.
  \end{minipage}
}
\makeatother

\begin{document}

\title{Spatially Consistent Air-to-Ground Channel Modeling and Simulation via 3D Shadow Projections}

\author{\IEEEauthorblockN{Evgenii Vinogradov$^{1,3}$, Aymen Fakhreddine$^{2}$, Abdul Saboor$^{3}$, Sergi Abadal$^{1}$, Sofie Pollin$^{3}$}
\IEEEauthorblockA{$^1$\textit{NaNoNetworking Center in Catalonia (N3Cat), Universitat Polit\`{e}cnica de Catalunya, Spain};
\\ $^2$\textit{Institute of Networked and Embedded Systems, University of Klagenfurt, Austria}
\\ $^3$\textit{Department of Electrical Engineering, KU Leuven, Belgium}; \\
Email: evgenii.vinogradov@upc.edu}
}

\maketitle
\begin{abstract}
We present an approach for spatially-consistent semi-deterministic Air-to-Ground (A2G) channel modeling in Unmanned Aerial Vehicle-assisted networks. We use efficient 3D building shadow projections to determine Line-of-Sight (LOS) regions, enabling fast generation of LOS maps. By integrating LOS-aware deterministic path loss with stochastic shadow fading, the approach produces spatially consistent A2G radio maps suitable for environment- and mobility-aware channel evaluation and performance prediction. Simulation results in ITU-compliant Manhattan grid environments demonstrate the model’s ability to reflect key urban propagation characteristics, such as LOS blockage patterns and outage behavior. The proposed approach provides an efficient alternative to ray tracing or fully stochastic models, with particular relevance for user mobility, link planning, and radio map generation in 6G non-terrestrial networks.
\end{abstract}


\section{Introduction}

Non-terrestrial networks, particularly those incorporating Unmanned Aerial Vehicles (UAVs) as Aerial Base Stations (ABS), are viewed as key enablers to achieve the ambitious coverage and flexibility goals of 6G~\cite{Akyildiz2020_6G}. The main advantage of UAV-based systems is their high probability of establishing Line-of-Sight (LOS) communication links with ground users~\cite{Vinogradov2018tut}. LOS links typically have lower path loss and a more stable signal compared to non-LOS (NLOS) links. Consequently, accurate modeling of LOS conditions is crucial for UAV-enabled communication systems, and numerous studies have addressed LOS probability modeling in various urban scenarios~\cite{Vinogradov2018tut,Saboor2024model,3GPPntn,saboorWCNC}.

In practical deployments, user mobility leads to transitions between LOS and NLOS conditions during a single communication session. Popular LOS probability models are not designed to reproduce these transitions and are unsuitable for dynamic scenarios~\cite{vinogradov2025prob}. In mobile scenarios, spatial consistency becomes critical. In terrestrial systems, the significance of spatial consistency has been well-documented~\cite{Karttunen2017spatial}, particularly for simulating cellular handovers. Ray-Tracing (RT) offers a solution, as it inherently preserves spatial consistency. However, RT’s computation time increases rapidly with growing scene complexity and trajectory length. To overcome this, semi-deterministic approaches~\cite{Colpaert2020handover,Li2021semideterministic, saboor2025cash} use RT solely for LOS determination, followed by statistical generation of other channel components. Alternatively, LOS regions can be defined using 3D building shadow projections~\cite{Kim2023features,cho2025placement}, enabling LOS estimation without point-by-point RT. 

\textit{State-of-the-art limitations:} RT-based methods are accurate but computationally demanding (even the semi-deterministic ones) for long trajectories or high-resolution simulations. Meanwhile, shadow projection methods have so far been limited to LOS probability estimation and have not been applied to spatially consistent channel generation.

\textit{Beyond state-of-the-art:} We propose a fast 3D building shadow projection approach (originally developed for real-time visual rendering) to deterministically construct spatially consistent LOS maps without RT. These maps are combined with LOS-aware channel models to produce A2G radio maps to predict how the channel evolve along user trajectories. This enables efficient simulation of user mobility effects, making our method suitable for UAV-based networks and environment-aware wireless optimization tasks such as ground user trajectory planning and coverage analysis. We implement this approach in the ITU-inspired Spatially Consistent A2G Channel simulator (ISCA2G), designed for urban grid environments.

The main contributions of this paper are:
\begin{itemize} 
\item Geometry-Based Shadow Projection (GBSP) approach for LOS map determination that uses shadows cast by 3D buildings. 
\item Spatially consistent A2G channel model that combines LOS-dependent deterministic path loss and stochastic shadow fading that evolves over space. 
\item Scalable simulator (ISCA2G) that integrates the above components and demonstrates their application in ITU-compliant urban grid environments to analyze channel behavior along a ground user trajectory.
\end{itemize}

The paper is organized as follows: Section~\ref{sec:system} presents the proposed GBSP approach for LOS map construction and the A2G channel model. Section~\ref{sec:Manhattan_sim} describes the implementation of the ISCA2G simulator. Section~\ref{sec:results} outlines the simulation setup and evaluates the spatial characteristics of the generated LOS segments and corresponding channels. Section~\ref{sec:conclusion} concludes the paper after a discussion of future research directions in Section~\ref{sec:future}.
\section{Spatially consistent A2G channel}\label{sec:system}

\begin{figure}
	\centering
	\includegraphics[width=\columnwidth]{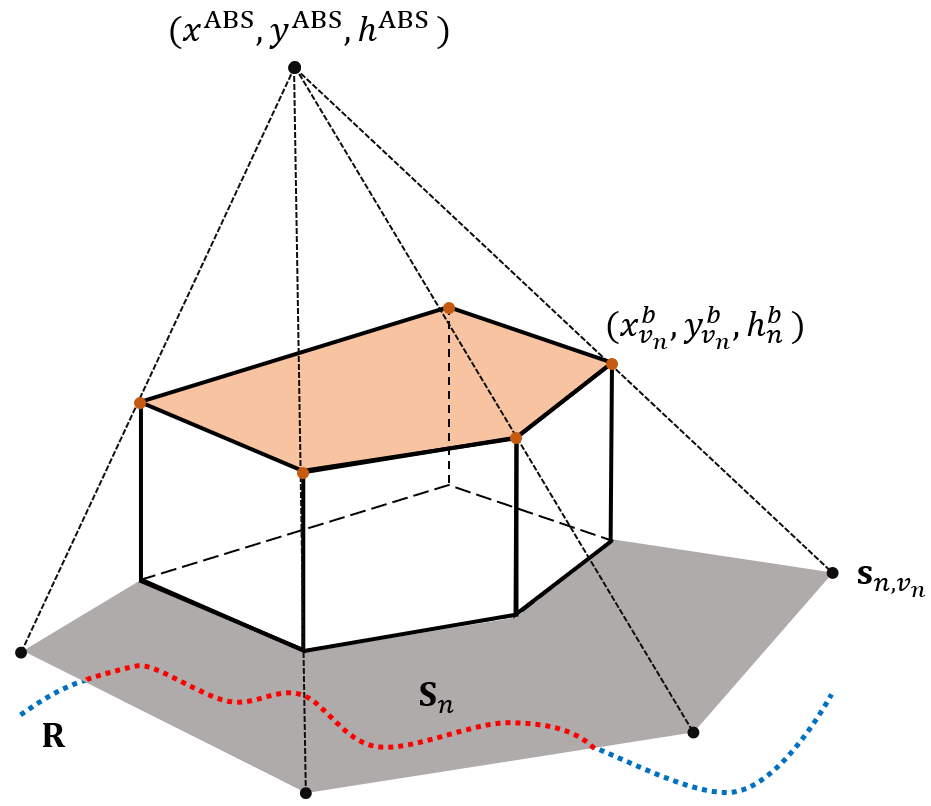}
	\caption{An example scenario of a system model depicting an $n-$th building described by $v_n=5$ vertices and the resulting shadow region $\mathbf{S}_n$. The dotted line indicates the user route $\mathbf{R}$ containing LOS and NLOS segments shown in blue and red, respectively.}
	\label{fig:system}

\end{figure}

To model spatially consistent A2G links, we consider a simplified yet representative urban environment consisting of a single ABS and a set of outdoor ground users. The ABS is located at a horizontal position $\mathbf{x}^{\text{ABS}} = (x^{\text{ABS}}, y^{\text{ABS}})$ and hovers at a height $h^{\text{ABS}}$. The environment includes a target area $A \subset \mathbb{R}^2$, which contains multiple buildings with arbitrary shapes that may obstruct the LOS link between the ABS and its users. The system is designed to capture how LOS conditions vary along the users' routes due to these urban obstructions as shown in Fig.~\ref{fig:system}. 

In this model, we assume that outdoor ground users are uniformly distributed over the ground area excluding building footprints, denoted by $\hat{A} \subset A$. This allows for a macroscopic analysis of LOS probability, where the LOS probability is determined by the portion of $\hat{A}$ not obstructed by building shadows. For a more detailed, user-specific analysis, we also consider the case of a single User Equipment (UE) moving along a predefined route $\mathbf{R}$\footnote{Our approach is agnostic to the way used to describe the route. Potentially, it can be defined as a function or a set of discrete waypoints $\mathbf{r}_{i} = (x^{\text{UE}}, y^{\text{UE}}) \in \hat{A}$ connected by lines.}. The route allows for determining the LOS conditions at each specific location, depending on the geometry of the environment. In this paper, we focus on the spatial aspect. The obtained results can be applied to scenarios with any user velocity or movement patterns, as the spatial model itself is independent of the specific dynamics of the user.

\subsection{Channel Model}
Assume communication between the ABS and a ground-based UE moving along a route. At any given location along the route, the channel losses (in dB) caused by large-scale effects can be calculated as:
\begin{equation}\label{eq:total_channel}
	\Lambda(\mathbf{r}_{i}) = \mathrm{\Lambda_0} + \mathrm{\Lambda_{ex}(\mathbf{r}_{i})} + \xi(\mathbf{r}_{i}),
\end{equation}
where \( \Lambda_0 \) is the free-space path loss (FSPL) at the reference distance \( d_0 \) and frequency \( f \), and \( \mathrm{\Lambda_{ex}(\mathbf{r}_{i})} \) is the excess path loss, which depends on the presence or absence of LOS and the geometry of the link between the ABS and the user location \( \mathbf{r}_i \). Additionally, \( \xi(\mathbf{r}_{i}) \) represents the zero-mean shadow fading, which has a LOS-dependent standard deviation $\sigma^\xi$. The shadow fading is spatially correlated with an autocorrelation function:
\begin{equation}\label{eq:ACF}
	R(\Delta d) = e^{-\frac{\Delta d}{d_{\text{decorr}}}},
\end{equation}
where $\Delta d$ is the distance between two points and $d_{\text{decorr}}$ is the decorrelation distance.

\subsection{Deterministic Geometry-Based Shadow Projection Approach for LOS modeling}\label{sec:gen_geometry}

Within the target area, we consider $N$ buildings, each modeled as a vertical prism with a polygonal base and a flat top. The base of the $n$-th building lies in the $xy$-plane and is represented as a simple polygon with $v_n$ vertices. Let $\mathbf{X}_{n}^b \in \mathbb{R}^{v_{n} \times 2}$ denote the coordinates of these base vertices and let $h_{n}^b$ denote the height of the building. We represent the building as the triple $\mathbf{b}_n = (v_n, \mathbf{X}_{n}^b, h_n^b)$, and define the set of all buildings as $\mathcal{B} = \{ \mathbf{b}_n \}_{n=1}^N$. The sides of each building form $v_n$ {wall faces}, each defined as a vertical rectangle connecting two consecutive base vertices along with their corresponding roof vertices at height $h_n^b$.

For each building $n$, we calculate the shadow of each roof vertex $v_n$ by projecting it onto the ground plane:
\begin{equation}\label{eq:shadow_vertex}
	\mathbf{s}_{n,v_n} = h^{\text{ABS}} \cdot \frac{(\mathbf{x}_{v_n}^b - \mathbf{x}^{\text{ABS}})}{h^{\text{ABS}} - h_{n}^b} + \mathbf{x}^{\text{ABS}}.
\end{equation}
{The shadow $\mathbf{S}_n$ cast by building $n$ is obtained as the union of (i) the building footprint, and (ii) shadows cast by all its wall faces.} The total shadowed region in the target area is the union of all individual building shadows:
\begin{equation}\label{eq:shadow_total}
	\mathbf{S}_{\text{total}} = \bigcup_{n} \mathbf{S}_n.
\end{equation}

\paragraph{LOS map definition and LOS probability}
The LOS map is defined as the subtraction of $\mathbf{S}_{\text{total}}$ from the outdoor area $\hat{A}$ as
\begin{equation}\label{eq:LOS map}
	\mathbf{L}= \hat{A} - \mathbf{S}_{\text{total}}.
\end{equation}
When ground users are uniformly distributed over $\hat{A}$, the NLOS probability can be calculated as the ratio of the shadowed subarea to $\hat{A}$, i.e.,
$$
P_{\text{NLOS}} = \frac{Area(\mathbf{S}_{\text{total}} \cap \hat{A})}{Area(\hat{A})},\quad P_{\text{LOS}}=1-P_{\text{NLOS}}.
$$

\paragraph{Route segmentation}
Finally, LOS and NLOS segments of the route $\mathbf{R}$ are defined as:
\begin{equation}\label{eq:NLOS}
	\mathbf{R}_{\text{LOS}} = \mathbf{R} \cap \mathbf{L},\quad \mathbf{R}_{\text{NLOS}} = \mathbf{R} \cap \mathbf{S}_{\text{total}} .
\end{equation}

\section{ISCA2G: ITU-Inspired Spatially Consistent A2G Channel Simulator}\label{sec:Manhattan_sim}

The proposed simulator ISCA2G\footnote{code is available on request via email to E. Vinogradov} applies the environment representation described in Section~\ref{sec:gen_geometry} to efficiently evaluate spatially consistent LOS and NLOS conditions for A2G links, considering a static ABS and a ground user following a piecewise linear route. The simulator generates urban layouts based on a Manhattan grid model~\cite{ITU} and determines LOS map by computing building shadows. Next, the LOS maps are used by the A2G channel model.

\subsection{LOS Simulation Workflow}
\begin{figure}
    \centering
    \includegraphics[width=\columnwidth]{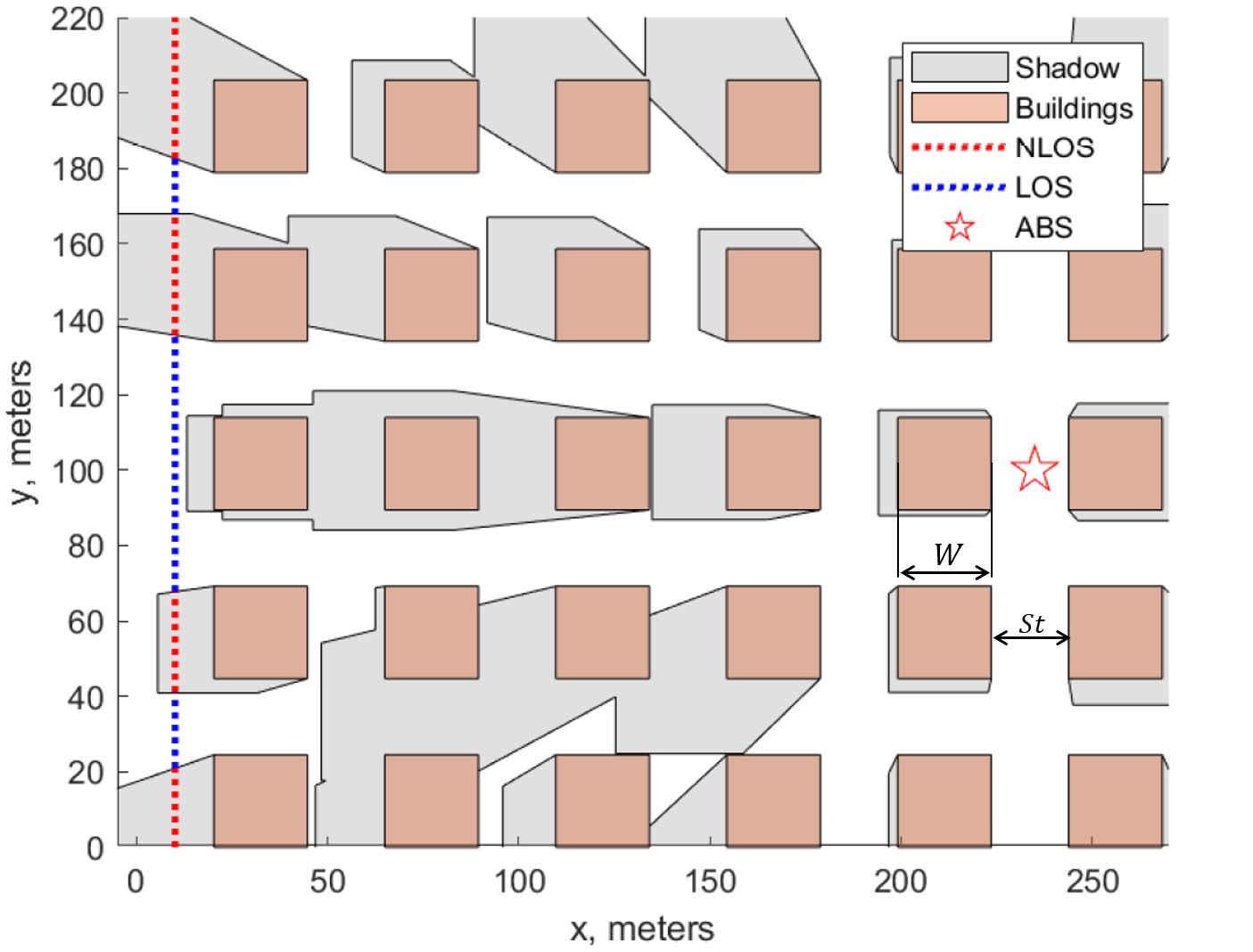}
    \caption{Example simulation (top view): ABS is deployed at 120 meters height in the Manhattan urban environment. The route is partly affected by shadows cast by square buildings placed as in~\cite{ITU}.}
    \label{fig:simulator_ex}
\end{figure}

ISCA2G follows a workflow to classify LOS/NLOS segments along a given route of the ground user:
\paragraph{Urban environment generation} Representative city layouts are generated using the environmental parameters $\alpha$, $\beta$, and $\gamma$, defined respectively as the build-up area ratio, building density per unit area, and the scale of the building height distribution, as specified by ITU~\cite{ITU}. Each realization consists of a grid where $N$ squared buildings of a width ${W}$ are separated by streets of width $St$ (see Fig.~\ref{fig:simulator_ex}). The widths are computed as:
\begin{align}
W = 1000 \sqrt{\frac{\alpha}{\beta}}, && St = \frac{1000}{\sqrt{\beta}} - W.
\end{align}
The building heights are randomly sampled from a Rayleigh distribution with scale parameter $\gamma$. The layout is composed of $N=I\cdot J$ buildings where $I$ and $J$ are the numbers of buildings along $x$ and $y$ axes, respectively.  

Each building $\mathbf{b}_{i,j}$ in $\mathcal{B}$ is represented by an ordered triple as $\mathbf{b}_{i,j} = (4, \mathbf{X}_{i,j}^b
, h_{i,j}^b)$ with the four vertices described by the building height and the rows of the matrix 
\begin{equation}
\mathbf{X}_4^b = \begin{bmatrix}
(i-1)\cdot(St+W)+St & (j-1)\cdot(St+W)\\
(i-1)\cdot(St+W)+St & (j-1)\cdot(St+W)+W\\
i\cdot(St+W) & (j-1)\cdot(St+W)\\
i\cdot(St+W) & (j-1)\cdot(St+W)+W
\end{bmatrix}    
\end{equation}

\paragraph{Route}
% $\frac{St}{2}$.
Similarly to previous simulators and models~\cite{Saboor2023simulator,Saboor2024model,Saboor2025pedestrian}, ISCA2G exploits the symmetry and homogeneity of the Manhattan layout: i) we focus on azimuth angles within the first quadrant $[0,\frac{\pi}{2}]$ and, ii) a complex route can be split into a number of lines connecting $L$ waypoints. In the simplest case, we consider a linear route described by two waypoints located in the middle of the street.
    
\paragraph{LOS map computation} For each building $\mathbf{b}_{i,j}$ in the grid, its shadow is computed by projecting all four roof vertices onto the ground using the geometric transformation in Eq. \eqref{eq:shadow_vertex}. The resulting set of projected points, combined with the base of the building, forms a shadow region defined as their convex hull. The total shadowed region $\mathbf{S}_{\text{total}}$ is obtained as the union of individual building shadows as in \eqref{eq:shadow_total} and used \eqref{eq:LOS map} to define the LOS map $\mathbf{L}$.
    
\paragraph{Route LOS/NLOS segmentation with LOS maps} The route of the mobile ground user $\mathbf{R}$ is tested for intersection with $\mathbf{L}$ and $\mathbf{S}_{\text{total}}$ using Eq.~\eqref{eq:NLOS} to define LOS and NLOS route segments, respectively. 
    
\subsection{ISCA2G Validation and Complexity Analysis}\label{sec:los_analysis}
We use RT simulations as a benchmark. The RT-based LOS/NLOS simulator (e.g., such as in our previous papers~\cite{Saboor2023simulator,vinogradov2025prob,saboorWCNC}) discretizes the user route into $M$ points. For each point $\mathbf{r}_i$, it casts a straight ray from the ABS to $\mathbf{r}_i$ and checks for intersections with all $N'$ buildings lying on the horizontal projection of the ray. If any intersection occurs below the building height, the point is marked NLOS; otherwise it is LOS. For simplicity of the following analysis, let us assume that all $N$ buildings can potentially cause NLOS for some part of the route. This yields a per-point LOS label via $M\times N$ intersection tests.

\paragraph{LOS verification} Both simulators produced identical LOS results for randomly generated locations of ABS and ground UEs, indicating the validity of the approach presented in Section~\ref{sec:gen_geometry} and ISCA2G in Section~\ref{sec:Manhattan_sim}. Additionally, both simulators reproduced the azimuth and elevation-dependent LOS probability estimation. Since the results are the same, we omit them for the sake of brevity. Note that the azimuth- and elevation-dependent behavior of LOS probability reproduced by ISCA2G and~\cite{Saboor2023simulator} can be modeled theoretically as demonstrated in~\cite{Saboor2024model,Saboor2025pedestrian}. Since the LOS probability model~\cite{Saboor2024model} shows good agreement with the 3GPP NTN model~\cite{3GPPntn}, we claim that the LOS probability results produced by ISCA2G are verified against the state-of-the-art models. While the theoretical models above are very useful and fast to compute, unfortunately, they do not account for spatial consistency. 
\paragraph{Complexity comparison}  
In the worst case, the RT-based method performs $\mathcal{O}(M \cdot N)$ building-ray intersection checks. Since each building has $v$ vertical walls, each intersection check is $\mathcal{O}(v)$, resulting in a total computational cost of $\mathcal{O}(M\,N\,v)$. Therefore, the runtime grows linearly with the sampling resolution $M$, the number of buildings $N$, and the fixed building shape complexity $v=4$.

By contrast, ISCA2G proceeds in three main steps:
\begin{enumerate}
	\item \emph{Shadow generation}: for each building, project its $v=4$ rooftop vertices to the ground, combine them with the $v=4$ base vertices, and compute their convex hull in $\mathcal{O}(1)$ time (since the total number of points is fixed at 8), for a total of $\mathcal{O}(N)$.
	\item \emph{Union of shadows}: merge the resulting $N$ simple polygons using a plane-sweep or divide-and-conquer algorithm in $\mathcal{O}(N \log N + K)$, where $K$ is the total number of edges in the final union polygon.
	\item \emph{Route intersection}: intersect the fixed polyline route (with $L-1$ linear segments defined by $L$ waypoints) against the unified shadow polygon in \(\mathcal{O}(L -1 + K)\).
\end{enumerate}
In our simulator, $v$ and $L$ are constants with a low value, hence, the dominant term is $\mathcal{O}(N \log N)$. The impact of $K$ can also be high, depending on the environment.

Thus, while RT-based LOS simulators such as~\cite{Saboor2023simulator} scale as $\mathcal{O}(MN)$ and become expensive for high resolution $M$, the shadow-based approach and ISCA2G have $\mathcal{O}(N\log N+K)$ complexity, offering computational savings for large-scale or high-resolution route analyses. Thus, this geometry-based method along with ISCA2G enables scalable, computationally efficient, and spatially consistent LOS and A2G channel modeling while circumventing the complexity of 3D RT techniques.

\subsection{LOS-aware A2G Channel Model}

\begin{table}[!t]
	\caption{Channel model parameters}\label{tab:channel}
	\begin{center}
		\begin{tabular}
			{c|c|c}
			Frequency & \multicolumn{2}{c}{2.5~GHz}\\
			$\Lambda_0$ & \multicolumn{2}{c}{Eq.~\eqref{eq:refPL}}\\
			$\rho$ & {LOS: 0.0272}& {NLOS: 2.3197}\\
			$\mu$ & {LOS: 0.7475}& {NLOS: 0.2361}\\
			$d_{\text{decorr}}$ & \multicolumn{2}{c}{11 m}\\

			%\hline

		\end{tabular}
	\end{center}
\end{table}
We adopt the channel model presented in Eq.~\eqref{eq:total_channel}, incorporating path loss and shadow fading formulations as in~\cite{Feng2006a2gPL}. {The spatial scales and types of urban environments considered in~\cite{Feng2006a2gPL} closely match the characteristics of our simulation setup.} The total path loss comprises two components: (i) the reference free-space attenuation depending on the ABS altitude and (ii) an elevation-angle-dependent excess loss. These terms together capture the 3D geometric nature of the A2G link. Similarly, shadow fading is parameterized using elevation-dependent standard deviations that differ for LOS and NLOS cases. To ensure the exponential shadow fading autocorrelation defined in Eq.~\eqref{eq:ACF}, we apply the well-known approach for spatially-consistent shadow fading map generation proposed by Claussen in~\cite{Claussen2005SFmap}.

The reference distance is defined as the relative height of the ABS directly above the UE (i.e., $d_0=h^{ABS}$ in our case). Hence
\begin{equation}\label{eq:refPL}
	\Lambda_0=20 \log_{10}\frac{{4\pi h^{\text{ABS}} f}}{c}. \end{equation}
As proposed in~\cite{Feng2006a2gPL}, the excess path loss depends solely on the elevation angle. In the LOS scenario, it is modeled as:
\begin{equation}
\Lambda_{\text{ex,LOS}}(\theta) = -20 \log_{10} \sin \theta,
\end{equation}
while for NLOS conditions, the excess path loss is
\begin{equation}
	\Lambda_{\text{ex,NLOS}}(\theta) = -16.16 + 12.0436 \exp(-\frac{90-\theta}{7.52}).
\end{equation}

Finally, the elevation-dependent shadow fading standard deviation for LOS and NLOS respectively is modeled as
\begin{equation}
    \sigma^\xi(\theta) = \rho (90-\theta)^\mu,
\end{equation}
where the parameters $\rho$ and $\mu$ are adopted from~\cite{Feng2006a2gPL} and can be found in Table~\ref{tab:channel} for the LOS and NLOS cases. The decorrelation distance is found to be varying between 9.5 and 12.9~m~\cite{Bucur2019LargeScale}, we use 11~m.

\section{Results}\label{sec:results}
We evaluate the proposed framework across four representative environments defined by ITU~\cite{ITU}: Suburban, Urban, Dense Urban, and High-Rise Urban. The corresponding environmental parameters, as well as other relevant simulation details, are summarized in Table~\ref{tab:setup}. These settings reflect progressively denser layouts with taller buildings.

Each simulation consists of 1000 independent realizations, where the UE follows a straight 1000-meter route. The ABS is randomly placed in each run, with horizontal coordinates drawn from $x, y \sim \mathcal{U}(0,1000)$~m and altitude $h^{\text{ABS}} \sim \mathcal{U}(30,250)$~m. 

The UE receiver sensitivity is --84.7~dBm~as per the 3GPP recommendation~\cite{3gppUE}. Strict Size-Weight-and-Power (SWAP) constraints for ABSs result in moderate transmit power levels and simple antennas with low gain. In our simulations, we assume an equivalent isotropically radiated power (EIRP) of 13, 18, and 23~dBm. Path loss and shadow fading are modeled as described in Section~\ref{sec:Manhattan_sim}. Outage is defined as the situation where the channels have a higher loss than $\Lambda_{\text{outage}}$ allowed by the receiver sensitivity and the transmitter EIRP: $\Lambda_{\text{outage}}=$EIRP -- UE sensitivity.

In the following, we focus on the spatial aspect of A2G communication, as the previous models~\cite{ITU,3GPPntn,Saboor2023simulator,Saboor2024model,Saboor2025pedestrian,Kim2023features,cho2025placement} sufficiently covered the LOS probability and statistical channel capacity analyses.

\begin{table}[!t]
\caption{Simulation parameters}\label{tab:setup}
\begin{center}
\begin{tabular}
{c|c|c|c|c}
&Suburban & Urban & Dense& High-Rise\\
&&&Urban&Urban \\ \hline
$\alpha$ & 0.1 & 0.3 & 0.5& 0.5 \\ 
$\beta$ & 750 & 500 & 300& 300 \\ 
$\gamma$ & 8 & 15 & 20& 50 \\ 
$St$ &25 & 20.2 & 16.9& 16.9 \\ 
$W$ &11.5 & 24.5 & 40.8& 40.8 \\ 
\hline
No. realizations & \multicolumn{4}{c}{1000}\\
ABS coordinates & \multicolumn{4}{c}{$x$ and $y \sim \mathcal{U}(0,1000)$~m, $h^{ABS} \sim \mathcal{U}(30,250)$~m}\\
Route length & \multicolumn{4}{c}{1000 m}\\
\hline
			UE sensitivity & \multicolumn{4}{c}{-84.7~dBm}\\
ABS EIRP & \multicolumn{4}{c}{23 dBm}\\
%\hline
\end{tabular}
\label{tab1}
\end{center}

\end{table}

\subsection{Spatially Consistent LOS Prediction}
\begin{figure}
    \centering
    \includegraphics[width=\columnwidth]{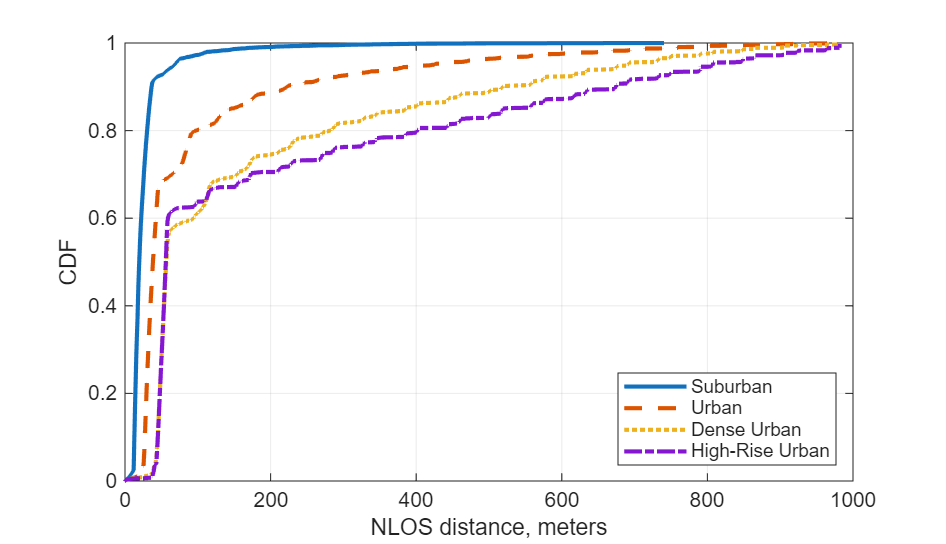}
    \includegraphics[width=\columnwidth]{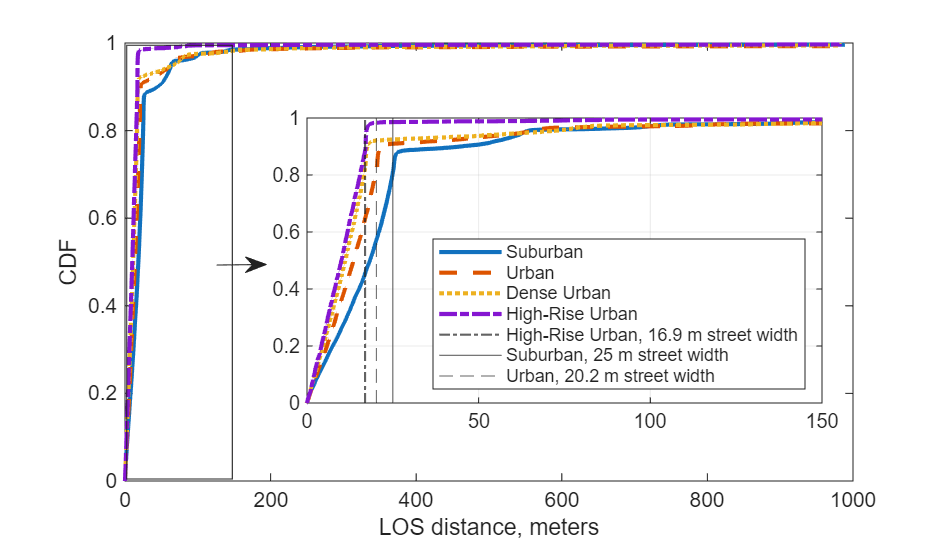}

        \caption{CDF of simulated NLOS (top) and LOS (bottom) distances in different environments. More sparse environments cause shorter NLOS and longer LOS segments. The shape of distributions is influenced by the regular layout and exhibits a relation to $W$ and $St$. }
    \label{fig:los_cdf}
\end{figure}
Fig.~\ref{fig:los_cdf} shows the distribution of the lengths of the NLOS and LOS segments in the four environments. The simulations demonstrate the effect of the regular Manhattan grid layout, where fixed building and street widths define the obstruction segments. For example, the Cumulative Distribution Function (CDF) of NLOS distances exhibits a stepped shape, with sharp increases occurring at regular intervals that are multiples of $W+St$ (i.e., a block of the Manhattan grid). Note that the NLOS segment length does not exceed one Suburban block in 90\% cases but this value decreases to 57-60\% in Dense and High-Rise Urban environments. On the other hand, over 80\% the LOS distances do not exceed the street width $St$.

\subsection{Channel and Outage}
\subsubsection{Outage probability}
\begin{figure}
    \centering
    \includegraphics[width=1\columnwidth]{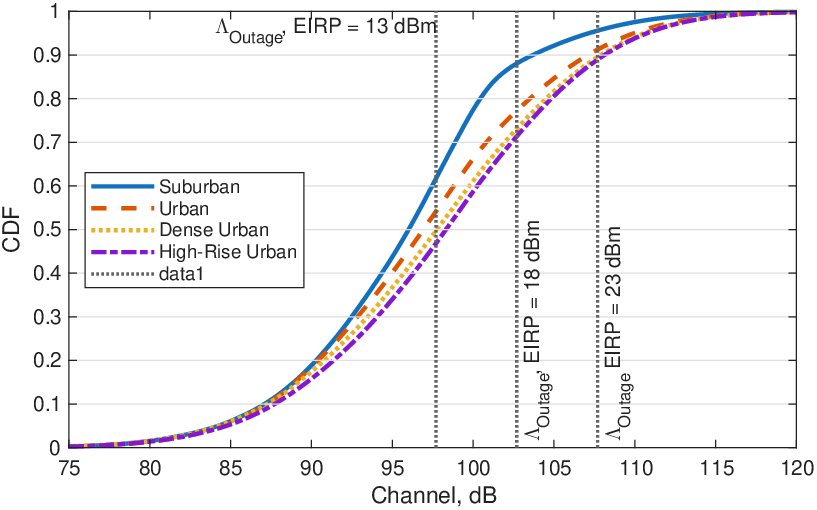}
    \caption{CDF of channel composed on path loss and shadow fading. Outage probability can be estimated based on the intersection of the CDFs with $\Lambda_{\text{outage}}$ allowed by the receiver sensitivity and the transmitter EIRP.}
    \label{fig:channel}

\end{figure}

Fig.~\ref{fig:channel} shows the distributions of channel attenuation obtained with ISCA2G for different environments. We observe that for EIRP = 23 dBm, outage probability does not exceed 11\% (in High-Rise Urban layout) and goes down to 4.3\% in Suburban environment. Since the route length is 1000~m, it is equivalent to 110 and 43~meters of outage. However, for ABS EIRP = 13 dBm, we observe 53.2\% and 38.4\% outage probability in the same setups (i.e., 532 and 384~m, respectively).

\subsubsection{Outage distances}
\begin{figure}
    \centering
    \includegraphics[width=\columnwidth]{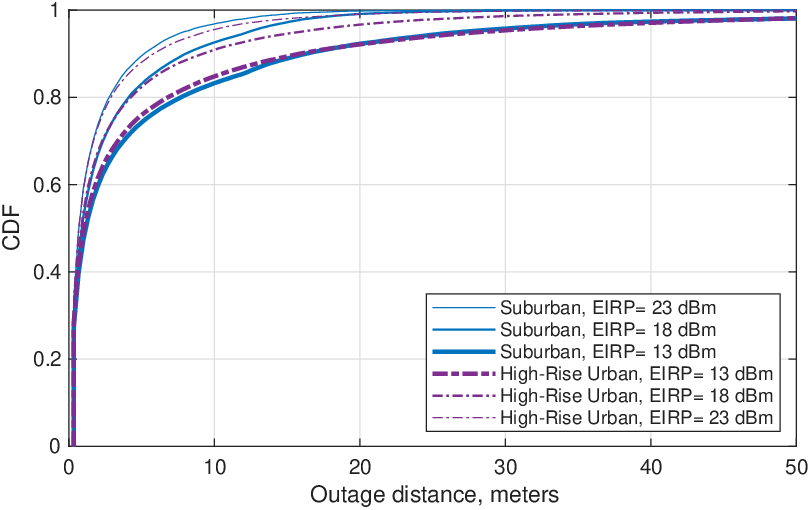}
    \caption{CDF of outage distances in Suburban and High-Rise Urban environments. The difference between the outage distances Urban, Dense Urban, and High-Rise Urban environments is negligible and we omit those curves.}
    \label{fig:outage}

\end{figure}

We define outage distance as segments of the UE route where the received signal power drops below the receiver sensitivity. The outage distance statistics for Suburban and High-Rise Urban environments are shown in Fig.~\ref{fig:outage}. The outage distances grow once EIRP is decreased, however, the distances do not exceed 9.6, 15.8, and 28 meters for EIRP = 23, 18, and 13 dBm with 95\% probability. Interestingly, the difference between the environments in terms of outage distances is not as significant as the difference in terms of the outage probability. Another observation is that the segments with the outage are relatively short (e.g., 28 m) compared to the total outage distance~(respectively, 532~m) indicating that a mobile user could experience unstable communication with short but frequent outages, especially with low-power ABSs deployed in High-Rise Urban environments. 

\section{Future Work}\label{sec:future}

Future work will extend the proposed framework toward full-stack spatially consistent Channel Knowledge Maps (CKMs) by (i) incorporating space and frequency consistent small-scale fading atop LOS maps and large scale channel components, (ii) generalizing to real-world layouts using GIS data (e.g., OpenStreetMap), (iii) parameterizing and validating A2G channel models via measurement campaigns\cite{Colpaert2024meas}, (iv) leveraging ML-based approaches for fast and consistent A2G CKM generation, (v) enabling UAV mobility, and (vi) integrating CKMs into system-level tasks such as beam planning, outage prediction, and mobility-aware access optimization.
 
\section{Conclusions}\label{sec:conclusion}
This paper presented a scalable approach for spatially consistent semi-deterministic air-to-ground channel modeling with deterministic LOS maps and path loss combined with stochastic shadow fading modeling. The approach projects 3D buildings into 2D shadows, enabling fast identification of LOS conditions and environment-aware transitions in channel characteristics, which are essential for mobile communication performance modeling. Note that the presented model formulation is applicable to arbitrary layouts, building shapes, and user trajectories.

Simulation results using regular ITU-defined Manhattan layouts confirm the model's ability to reflect the effects of the urban environment. In particular, the NLOS distance CDFs show stepped increases tied to multiples of block sizes (consisting of a building and streets), while over 80\% of LOS segments are shorter than the street width. Channel simulations show that outage probability remains below 11\% for high ABS power (EIRP = 23~dBm) but increases sharply to over 50\% for low-power ABSs (EIRP = 13~dBm). Meanwhile, outage distance remains modest (under 28~m with 95\% probability), even in dense urban settings and low EIRP of 13 dBm.
\section*{Acknowledgments}
This work is part of the I+D+i project titled BLOSSOMS, grant PID2024-158530OB-I00, funded by MICIU/AEI/10.13039/501100011033/ and by ERDF/EU. Aymen Fakhreddine's contribution is funded by the Austrian Science Fund (FWF\,--\,Der Wissenschaftsfonds) under grant ESPRIT-54 (Grant DOI: 10.55776/ESP54).

\bibliographystyle{IEEEtran}
\bibliography{main}

@INPROCEEDINGS{vinogradov2025prob,
  author={Vinogradov, Evgenii and Saboor, Abdul and Cui, Zhuangzhuang and Fakhreddine, Aymen},
  booktitle={2025 IEEE 101st Vehicular Technology Conference (VTC2025-Spring)}, 
  title={Spatially Consistent Air-to-Ground Channel Modeling with Probabilistic {LOS/NLOS} Segmentation}, 
  year={2025},
  volume={},
  number={},
  pages={1-5},
  doi={10.1109/VTC2025-Spring65109.2025.11174559}}

@ARTICLE{Saboor2024model,
  author={Saboor, Abdul and Vinogradov, Evgenii and Cui, Zhuangzhuang and Al-Hourani, Akram and Pollin, Sofie},
  journal={IEEE Open Journal of the Communications Society}, 
  title={{A Geometry-Based Modelling Approach for the Line-of-Sight Probability in UAV Communications}}, 
  year={2024},
  volume={5},
  number={},
  pages={364-378},
  doi={10.1109/OJCOMS.2023.3341627}}

@ARTICLE{Saboor2025pedestrian,
  author={Saboor, Abdul and Cui, Zhuangzhuang and Vinogradov, Evgenii and Pollin, Sofie},
  journal={IEEE Antennas and Wireless Propagation Letters}, 
  title={{Air-to-Ground Channel Model for Pedestrian and Vehicle Users in General Urban Environments}}, 
  year={2025},
  volume={24},
  number={1},
  pages={227-231},
  doi={10.1109/LAWP.2024.3492507}}

@INPROCEEDINGS{Saboor2023simulator,
  author={Saboor, Abdul and Vinogradov, Evgenii and Cui, Zhuangzhuang and Pollin, Sofie},
  booktitle={IEEE International Black Sea Conference on Communications and Networking (BlackSeaCom)}, 
  title={{Probability of Line of Sight Evaluation in Urban Environments using 3D Simulator}}, 
  year={2023},
  volume={},
  number={},
  pages={135-140},
  doi={10.1109/BlackSeaCom58138.2023.10299705}}

@INPROCEEDINGS{Bucur2019LargeScale,
  author={Bucur, Madalina and Sorensen, Troels and Amorim, Rafhael and Lopez, Melisa and Kovacs, Istvan Z. and Mogensen, Preben},
  booktitle={2019 IEEE 90th Vehicular Technology Conference (VTC2019-Fall)}, 
  title={Validation of Large-Scale Propagation Characteristics for {UAVs} within Urban Environment}, 
  year={2019},
  volume={},
  number={},
  pages={1-6},
  doi={10.1109/VTCFall.2019.8891422}}

@article{3gppUE,
  title={{User Equipment (UE) radio transmission and reception, document 38.101}},
  journal={{3rd Generation Partnership Project, Technical Specification Group Radio Access Network}},
  year={2018}
}

@INPROCEEDINGS{Feng2006a2gPL,
  author={Qixing Feng and McGeehan, J. and Tameh, E.K. and Nix, A.R.},
  booktitle={IEEE 63rd Vehicular Technology Conference}, 
  title={Path Loss Models for Air-to-Ground Radio Channels in Urban Environments}, 
  year={2006},
  volume={6},
  number={},
  pages={2901-2905},
  doi={10.1109/VETECS.2006.1683399}}

@INPROCEEDINGS{saboorWCNC,
  author={Saboor, Abdul and Cui, Zhuangzhuang and Vinogradov, Evgenii and Pollin, Sofie},
  booktitle={2025 IEEE Wireless Communications and Networking Conference (WCNC)}, 
  title={{Empirical Line-of-Sight Probability Modeling for UAVs in Random Urban Layouts}}, 
  year={2025},
  volume={},
  number={},
  pages={1-6},
  doi={10.1109/WCNC61545.2025.10978620}}

@ARTICLE{Karttunen2017spatial,
  author={Karttunen, Aki and Molisch, Andreas F. and Hur, Sooyoung and Park, Jeongho and Zhang, Charlie Jianzhong},
  journal={IEEE Transactions on Wireless Communications}, 
  title={{Spatially Consistent Street-by-Street Path Loss Model for 28-GHz Channels in Micro Cell Urban Environments}}, 
  year={2017},
  volume={16},
  number={11},
  pages={7538-7550},
  doi={10.1109/TWC.2017.2749570}}

@ARTICLE{cho2025placement,
 author={Cho, Yeonwoo and Won, Jonghyeon and Kim, Do-Yup and Lee, Jang-Won},
  journal={IEEE Internet of Things Journal}, 
  title={Optimal Placement of Aerial Base Station Utilizing Topographic Features}, 
  year={2025},
  volume={12},
  number={12},
  pages={19882-19900},
  doi={10.1109/JIOT.2025.3545125}}

@ARTICLE{Kim2023features,
  author={Kim, Do-Yup and Saad, Walid and Lee, Jang-Won},
  journal={IEEE Transactions on Wireless Communications}, 
  title={On the Use of High-Rise Topographic Features for Optimal Aerial Base Station Placement}, 
  year={2023},
  volume={22},
  number={3},
  pages={1868-1884},
  doi={10.1109/TWC.2022.3207427}}

@article{ITU,
  title={Propagation data and prediction methods required for the design of terrestrial broadband radio access systems operating in a frequency range from 3 to 60 {GHz},},
  author={Series, P},
  journal={Recommendation ITU-R P.1410-6},
  year={2023},
}

@INPROCEEDINGS{Li2021semideterministic,
  author={Li, Yuanbo and Li, Ning and Han, Chong},
  booktitle={ICC 2021 - IEEE International Conference on Communications}, 
  title={{Ray-tracing Simulation and Hybrid Channel Modeling for Low-Terahertz UAV Communications}}, 
  year={2021},
  volume={},
  number={},
  pages={1-6},
  doi={10.1109/ICC42927.2021.9500549}}

@ARTICLE{Akyildiz2020_6G,
  author={Akyildiz, Ian F. and Kak, Ahan and Nie, Shuai},
  journal={IEEE Access}, 
  title={{6G and Beyond: The Future of Wireless Communications Systems}}, 
  year={2020},
  volume={8},
  number={},
  pages={133995-134030},
  doi={10.1109/ACCESS.2020.3010896}}

@article{Vinogradov2018tut,
  title = {Tutorial on {{UAVs}}: A Blue Sky View on Wireless Communication},
  shorttitle = {Tutorial on {{UAVs}}},
  author = {Vinogradov, Evgenii and Sallouha, Hazem and De~Bast, SIbren and Azari, Mahdi and Pollin, Sofie},
  date = {20180101},
  journal = {Journal of Mobile Multimedia},
  volume = {14},
  pages = {395--395},
  publisher = {{River Publishers}},
  issn = {1550-4654},
  doi = {10.13052/jmm1550-4646.1443},
  number = {4},
  year={2018},
}

@article{3GPPntn,
  title={{Study on New Radio (NR) to Support non terrestrial networks}},
  journal={{3rd Generation Partnership Project, Technical specification group radio access network, 3GPP, document TR 38.811}},
  year={2018}
}

@INPROCEEDINGS{Colpaert2020handover,
  author={Colpaert, Achiel and Vinogradov, Evgenii and Pollin, Sofie},
  booktitle={2020 IEEE Globecom Workshops}, 
  title={{3D beamforming and handover analysis for UAV networks}}, 
  year={2020},
  volume={},
  number={},
  pages={1-6},
  doi={10.1109/GCWkshps50303.2020.9367570}}

@ARTICLE{Colpaert2024meas,
  author={Colpaert, Achiel and Cui, Zhuangzhuang and Vinogradov, Evgenii and Pollin, Sofie},
  journal={IEEE Transactions on Vehicular Technology}, 
  title={{3D Non-Stationary Channel Measurement and Analysis for MaMIMO-UAV Communications}}, 
  year={2024},
  volume={73},
  number={5},
  pages={6061-6072},
  doi={10.1109/TVT.2023.3340447}}

@INPROCEEDINGS{Claussen2005SFmap,
  author={Claussen},
  booktitle={2005 IEEE 16th International Symposium on Personal, Indoor and Mobile Radio Communications}, 
  title={Efficient modelling of channel maps with correlated shadow fading in mobile radio systems}, 
  year={2005},
  volume={1},
  number={},
  pages={512-516},
  doi={10.1109/PIMRC.2005.1651489}}

@INPROCEEDINGS{saboor2025cash,
      title={{CASH: Context-Aware Smart Handover for Reliable UAV Connectivity on Aerial Corridors}}, 
      author={Abdul Saboor and Zhuangzhuang Cui and Achiel Colpaert and Evgenii Vinogradov and Sofie Pollin},
      year={2025},
    booktitle={To appear in IEEE Global Communications Conference},
      eprint={2508.03862},
      archivePrefix={arXiv},
      primaryClass={cs.NI},
      url={https://arxiv.org/abs/2508.03862}, 
}

\end{document}